\documentclass[pra,aps,10pt,superscriptaddress,twocolumn,floatfix,showpacs]{revtex4-1}
\usepackage{graphicx}
\usepackage{amsfonts}
\usepackage{epsfig}
\usepackage[pdftex]{color}
\usepackage{epstopdf}
\usepackage{amsmath,graphicx,amssymb,braket,subfigure,upgreek}
\usepackage[colorlinks=true, linkcolor=blue, citecolor=blue, urlcolor=blue, breaklinks=true]{hyperref}
\usepackage{bigints}
\usepackage{dcolumn}
\usepackage{bm}
\usepackage{mathtools}
\usepackage{mathrsfs}
\usepackage{frcursive}
\usepackage{dsfont}
\usepackage{bbm}

\newcommand{\mi}{{\rm i}}

\begin{document}
\title{
Effects of energy extensivity on the quantum phases of long-range interacting systems
}

\author{T.~Botzung}
\affiliation{ISIS (UMR 7006) and icFRC, University of Strasbourg and CNRS, 67000 Strasbourg, France}
\author{D.~Hagenm\"uller}
\thanks{dhagenmuller@unistra.fr}
\affiliation{ISIS (UMR 7006) and icFRC, University of Strasbourg and CNRS, 67000 Strasbourg, France}
\author{G.~Masella}
\affiliation{ISIS (UMR 7006) and icFRC, University of Strasbourg and CNRS, 67000 Strasbourg, France}
\author{J.~Dubail}
\affiliation{LPCT (UMR7019), Universit\'{e} de Lorraine and CNRS, F-54506 Vandoeuvre-les-Nancy, France}
\author{N.~Defenu}
\affiliation{Institute for Theoretical Physics, Heidelberg University, D-69120 Heidelberg, Germany}
\author{A.~Trombettoni}
\affiliation{SISSA and INFN, Sezione di Trieste, I-34136
Trieste, Italy}
\author{G.~Pupillo}
\thanks{pupillo@unistra.fr}
\affiliation{ISIS (UMR 7006) and icFRC, University of Strasbourg and CNRS, 67000 Strasbourg, France}

\date{\today}

\begin{abstract}
We investigate the ground state properties of one-dimensional hard-core bosons interacting via a variable long-range potential using the density matrix renormalization group. We demonstrate that restoring energy extensivity in the system, which is done by rescaling the interaction potential with a suitable size-dependent factor known as Kac's prescription, has a profound influence on the low-energy properties in the thermodynamic limit. While an insulating phase is found in the absence of Kac's rescaling, the latter leads to a new metallic phase that does not fall into the conventional Luttinger liquid paradigm. Our results in one dimension illustrate the subtlety of determining the thermodynamic properties of generic long-range quantum systems.
\end{abstract}

\maketitle

Long-range (LR) interacting systems are characterized by highly non-local couplings between their constituents, typically decaying as a power law for large distances between them. LR models feature a wide variety of applications including self-gravitating clusters~\cite{Padmanabhan1990}, ferromagnetic materials~\cite{Landau1960}, non-neutral plasmas~\cite{Nicholson1983}, cavity-QED systems~\cite{Ritsch2013}, and 1D quantum wires~\cite{Giamarchi2004}. Recent progress in the realization of artificial lattices of cold gases with sizable LR interactions has stimulated considerable interest~\cite{Stuhler2005,Ni2008,Carr2009,Lahaye2009,Schneider2012,Britton2012,Jin2012,Bermudez2013,Richerme2014,Jurcevic2014,Douglas2015,Baier2016,Schmitt2016,Chomaz2018,Vaidya2018,Lepoutre2018,Tang2018a}. In parallel, theoretical studies of the ground state of LR spin models have revealed anomalous critical exponents~\cite{Defenu2016,Defenu2017,Defenu2019_1} and decay of correlations~\cite{Dutta2001,Laflorencie2005,Peter2012,Hauke2010,Frerot2017,Botzung2019}, as well as the existence of new quantum phases~\cite{Burnell2009,Pollet2010,Capogrosso2010,Vodola2014,Maghrebi2017,Masella2019}.

An archetypal and still actively studied LR quantum model consists of one-dimensional (1D) fermions interacting via a $1/r$ (unscreened) Coulomb potential, with $r$ the distance separating two particles. Schulz showed using bosonization techniques that the ground state of this system is a peculiar metal resembling a classical Wigner crystal, with very slow decay of the charge correlations associated to the plasmon mode~\cite{Schulz1993,Wang2001}. This result was confirmed numerically using density matrix renormalization group (DMRG)~\cite{Fano1999} and variational Monte Carlo methods~\cite{Casula2006,Astrakharchik2011,Lee2011}. In the presence of a lattice at commensurate fillings, it was shown that while the metallic behavior is surprisingly enhanced as compared to short-range interactions for small system size, the ground state ultimately enters an insulating phase in the thermodynamic limit~\cite{Poilblanc1997,Capponi2000,Valenzuela2003,Li2019}.

The {\em strong} LR regime for a $d$-dimensional system with volume $\mathcal{V}$ is achieved when the power-law exponent $\alpha$ entering the potential $V(r) \propto 1/r^\alpha$ is such that $0\leq \alpha \leq d$. This regime is typically associated to unusual thermodynamic properties such as a non-extensive energy $E\sim \mathcal{V}^{2-\frac{\alpha}{d}}$ leading to an ill-defined thermodynamic limit~\cite{Ruelle1963}. Furthermore, the total energy cannot be obtained by summing up the energies of different subsystems as is usually the case for short-range interactions~\cite{Dauxois2002,Mukamel2008}. This non-additivity appears as a fundamental property of LR models and leads to exotic behaviors including the breaking of ergodicity, the existence of slow relaxation processes, and the inequivalence of statistical ensembles~\cite{Barre2001,Campa2009,Kastner2010,Levin2014}. In contrast, extensivity can be restored by rescaling the interaction potential with an appropriate volume-dependent factor $\Lambda$, which is known as Kac's prescription~\cite{Kac1963}. The latter is systematically used to study the thermodynamic properties of classical spin models with LR interactions~\cite{Cannas1996,Tamarit2000,Dauxois2002_chapter,Campa2003,Kastner2006}, where the dynamical properties with and without Kac's rescaling are the same provided the respective time scales $t_{\rm res}$ and $t$ satisfy $t=t_{\rm res}/\sqrt{\Lambda}$~\cite{Anteneodo1998,Anteneodo2004}. However, it can be shown that the latter statement does not hold true in quantum systems. It is an open and interesting question to investigate what other properties such as ground state phases can be fundamentally modified by Kac's rescaling.

Here we study a 1D periodic chain of hard-core bosons in the strong LR regime at half-filling, using the Luttinger liquid (LL) theory combined with DMRG calculations~\cite{ITensor} for large system sizes ($\gtrsim 200$ sites). In the absence of Kac's rescaling, we find that the ground state consists of an insulating gapped phase in the thermodynamic limit in the whole range $0 < \alpha \leq 1$, extending the results of Ref.~\cite{Capponi2000} in the marginal case $\alpha=1$. In stark contrast, we demonstrate that Kac's rescaling leads to a metallic phase for any finite strength of the interaction in the thermodynamic limit for $0 \leq \alpha \leq 1$. This finding raises fundamental questions on how to study the thermodynamics of LR interacting quantum systems. Surprisingly, this metallic phase is found to be incompatible with a conventional LL, which is demonstrated by computing the Luttinger parameters from the single-particle correlation function, the structure factor, the charge gap and the charge stiffness. Restoring extensivity is further shown to eliminate the plasmon modes while preserving the LR character of the potential, and with it some inherent properties of the strong LR regime such as non-addivity. Physical realizations of our model may be obtained in the context of cavity-QED with cold atoms~\cite{Ritsch2013}.


We consider the Hamiltonian 
\begin{align}
H= -t \sum_{i=1}^{L} \left(a^{\dagger}_i  a_{i+1}  + {\rm h.c.} \right) + \sum_{i>j} V^{(\alpha)}_{i-j} n_{i} n_{j},
\label{full_H}
\end{align}
where the operator $a_i$ ($a_i^{\dagger}$) annihilates (creates) a hard-core boson on site $i=1,\cdots,L$, and $n_i = a_i^{\dagger} a_i$ is the local density. The interaction potential reads
\begin{align*}
V^{(\alpha)}_{i-j}=\frac{V}{\Lambda_{\alpha} (L) r_{i-j}^{\alpha}}  \qquad V>0,
\end{align*}
where $r_{i-j}=(La/\pi)\sin(\pi \vert i- j \vert/L)$ for $L\gg 1$ as we assume periodic boundary conditions (circular chain~\cite{Note1}). In the following, the nearest neighbor hopping $t$ and lattice spacing $a$ are set to $t \equiv a \equiv 1$. Kac's rescaling of the interaction potential is included via the function $\Lambda_{\alpha} (L)=1$ for $\alpha > 1$ (absence of Kac's rescaling), $\Lambda_{\alpha}(L)=\log(L)$ in the marginal case $\alpha= 1$, and $\Lambda_{\alpha}(L)=L^{1-\alpha}$ in the strong LR regime $\alpha < 1$. Note that the XXZ Heisenberg model with LR coupling along the $z$ direction and short-range coupling along $x$ and $y$ can be mapped onto hard-core bosons Eq.~(\ref{full_H}) or spinless fermions via a Jordan-Wigner transformation.

\begin{figure}[ht]
\includegraphics[width=\columnwidth]{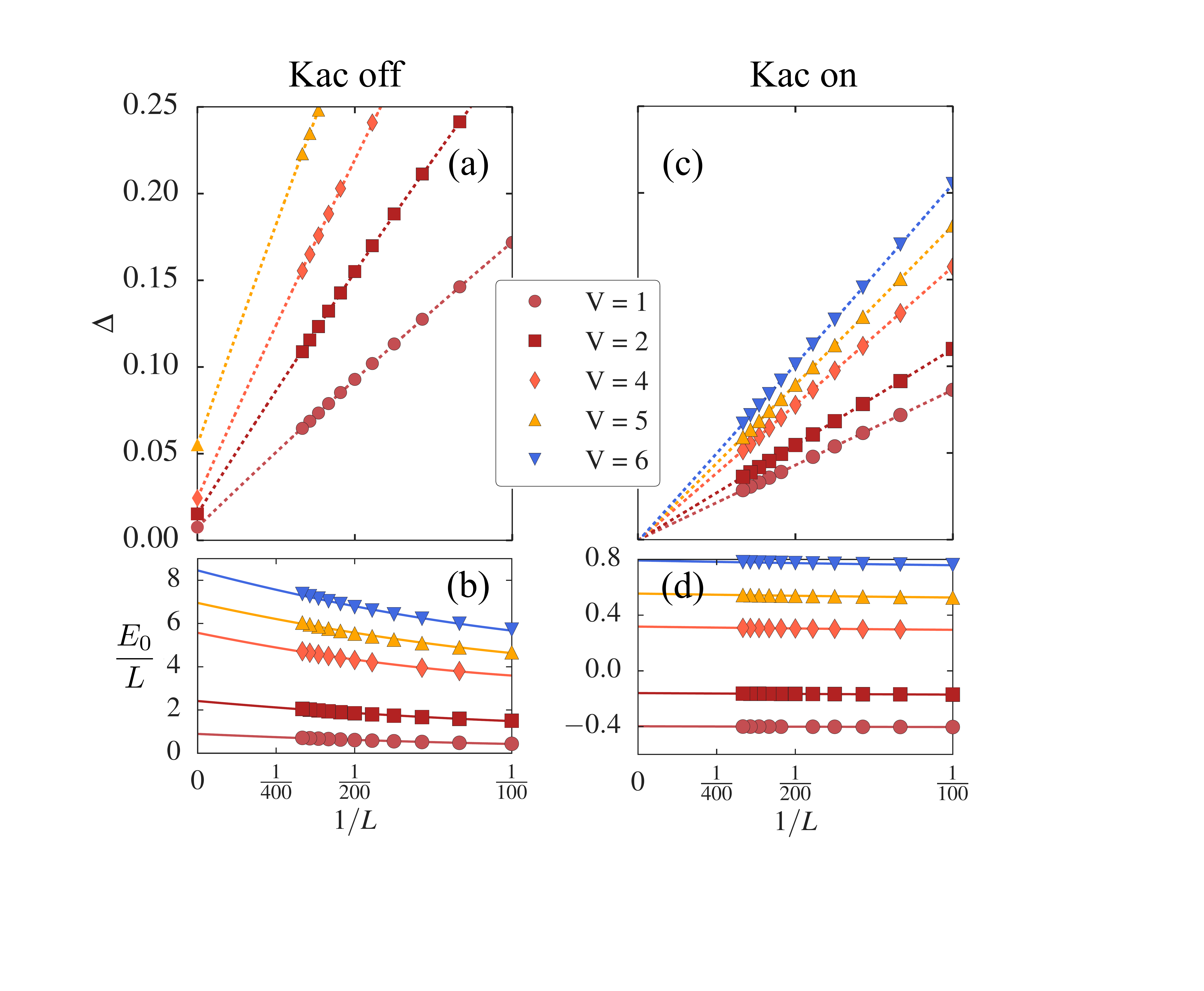}
\caption{Finite-size scaling of the single-particle gap $\Delta$ computed with DMRG at half-filling $\langle n_{i} \rangle =0.5$, for $\alpha = 1$ and different interaction strengths $V$ (in units of the hopping energy $t$). While an insulating phase ($\Delta\neq 0$) is found without Kac's rescaling \textbf{(a)} \& \textbf{(b)}, the latter leads to a metallic phase ($\Delta =0$) in the thermodynamic limit \textbf{(c)} \& \textbf{(d)}. Extrapolation for $L\to \infty$ is obtained by fitting the numerical data with $\Delta (L) = b +  \frac{c}{L} + \frac{d}{L^{2}}$ (dotted lines). The finite-size scaling of the ground state energy per particle is shown in \textbf{(b)} \& \textbf{(d)}.}
\label{Fig1}
\end{figure}

The main result of our work is summarized in Fig.~\ref{Fig1}, where we compute the single-particle charge gap $\Delta=E_{0} (N+1)+E_{0} (N-1)-2 E_{0} (N)$ for $\alpha=1$ and different interaction strengths $V$. Here, $E_{0} (N)$ denotes the energy of the ground state with $N$ particles. The situation without Kac's rescaling [$\Lambda_{1}(L)=1$, Fig.~\ref{Fig1} \textbf{(a)} \& \textbf{(b)}] has been already investigated in Ref.~\cite{Capponi2000}, and features a non-extensive energy. In this case, we find that the gap $\Delta (L\gg 1)\neq 0$ for any $V> 0$, which indicates an insulating phase in the thermodynamic limit consistently with the conclusion of Ref.~\cite{Capponi2000}. These results are drastically modified when using the Kac's prescription. By rescaling the interaction potential [$\Lambda_{1}(L)=\log (L)$, Fig.~\ref{Fig1} \textbf{(c)} \& \textbf{(d)}], we find that while extensivity is clearly restored, $\Delta\sim 1/L$ for all $V>0$. This result indicates a metallic behavior in the thermodynamic limit observed in the whole range $0\leq \alpha \leq 1$ (not shown).   

In order to understand the physical origin of the significant difference between the extensive and non-extensive models, we investigate the low-energy properties of Eq.~(\ref{full_H}) using the LL theory. A convenient bosonic representation of $H$ in terms of the continuous variable $x\equiv j a$ can be obtained by treating the interaction potential as a perturbation~\cite{Giamarchi2004}  
\begin{align}
H = \frac{1}{2\pi}\int \!\! dx \; u K \left(\pi \Pi \right)^{2} + \frac{u}{K} \nabla^{2} \phi  - \frac{g}{\pi a^{2}} \cos \left(4 \phi \right),
\label{cont_H}
\end{align}
where $\Pi (x)$ and $\phi (x)$ and are canonically conjugate bosonic fields depending on the long wavelength fluctuations of the fermion density. The so-called Luttinger parameters $u$ and $K$ are related by~\cite{Capponi2000,Note1}
\begin{align}
uK &= v_{\rm F} \nonumber \\ 
\frac{u}{K} &= v_{\rm F} + \frac{1}{\pi} \sum_{r=1}^{L} V^{(\alpha)}_{r} \left[1-\cos\left(2 k_{\rm F} r\right) \right] 
\label{eq: LL_parameters}
\end{align}
where $v_{\rm F}$ denotes the Fermi velocity and $k_{\rm F}$ the Fermi wave vector. The first two (quadratic) terms of Eq.~(\ref{cont_H}) describe how the properties of the non-interacting LL are renormalized by the interactions. In particular, $K$ determines the decay of the single-particle correlation function $\langle a_i^{\dagger} a_j \rangle \sim r_{i-j}^{-1/(2K)}$. The third term in Eq.~(\ref{cont_H}) stems from scattering processes across the Fermi surface where the particle momentum is conserved up to a reciprocal lattice vector. It is usually denoted as umklapp term and scales with the strength 
\begin{equation}
g = \sum_{r=1}^{L} V^{(\alpha)}_{r} \cos\left(2 k_{\rm F} r\right).
\label{eq: umklapp}
\end{equation}
For a finite $g$, it is possible to show using a renormalization-group study~\cite{Giamarchi2004} of the Hamiltonian Eq.~(\ref{cont_H}) that the system goes from an insulating to a metallic phase as $K$ is increased above a critical value $K_c$. At half-filling, and neglecting multiple umklapp scattering~\cite{Schmitteckert2004}, the critical value is $K_c= 0.5$. Note that in the case of a nearest-neighbor interaction $\alpha \to \infty$, such a metal-insulator transition occurs at $V=2t$~\cite{Franchini2017}. 

We consider a half-filled band $\langle n_{i} \rangle=0.5$, which provides $k_{\rm F}=\pi/2$ and $v_{\rm F}=2$. In the absence of Kac's rescaling, the first sum $\sum_{r} V^{(\alpha)}_{r}$ entering Eq.~(\ref{eq: LL_parameters}) diverges in the thermodynamic limit $\sim \log (L)$ for $\alpha=1$ and $\sim L^{1-\alpha}/(1-\alpha)$ for $0\leq\alpha<1$. The second sum $\sum_{r} V^{(\alpha)}_{r} \cos\left(2 k_{\rm F} r\right)$ entering Eqs.~(\ref{eq: LL_parameters}) and (\ref{eq: umklapp}) is bounded due to the alternating sign. Therefore, while the umklapp scattering strength $g$ remains finite, the Luttinger parameter $K \to 0$ for $0 <\alpha \leq 1$ and $V >0$ in the thermodynamic limit, consistently with an insulating phase. 

We find that rescaling the interaction potential with the factor $\Lambda_{\alpha} (L)= \log(L)$ for $\alpha=1$ and $\Lambda_{\alpha} (L)= L^{1-\alpha}$ for $\alpha < 1$ strongly affects the competition between $K$ and $g$. In this case, the long-wavelength divergence is removed since $\lim_{L \to \infty} \sum_{r} V^{(\alpha)}_{r} = V$ for $\alpha=1$ and $\lim_{L \to \infty} \sum_{r} V^{(\alpha)}_{r} = V/(1-\alpha)$ for $\alpha<1$. This suggests a metallic phase for $0 < \alpha \leq 1$, since $K$ remains finite and $g \to 0$ for any finite $V >0$ in the thermodynamic limit as seen from Eqs.~(\ref{eq: LL_parameters}) and (\ref{eq: umklapp}). 

The above arguments cannot be used in the case $\alpha=0$ since the series $\sum_{r} V^{(\alpha)}_{r} \cos\left(2 k_{\rm F} r\right)$ does not have a unique limit for $L\to \infty$. Nevertheless, this particular case can be solved exactly using a mean-field approach. This leads to a free fermion (metallic) phase with charge correlations $\langle a_i^{\dagger} a_j \rangle \sim r_{ij}^{-1/2}$, regardless of the presence or absence of Kac's rescaling~\cite{Note1,Note3}.  
\begin{figure}[ht]
\includegraphics[width=\columnwidth]{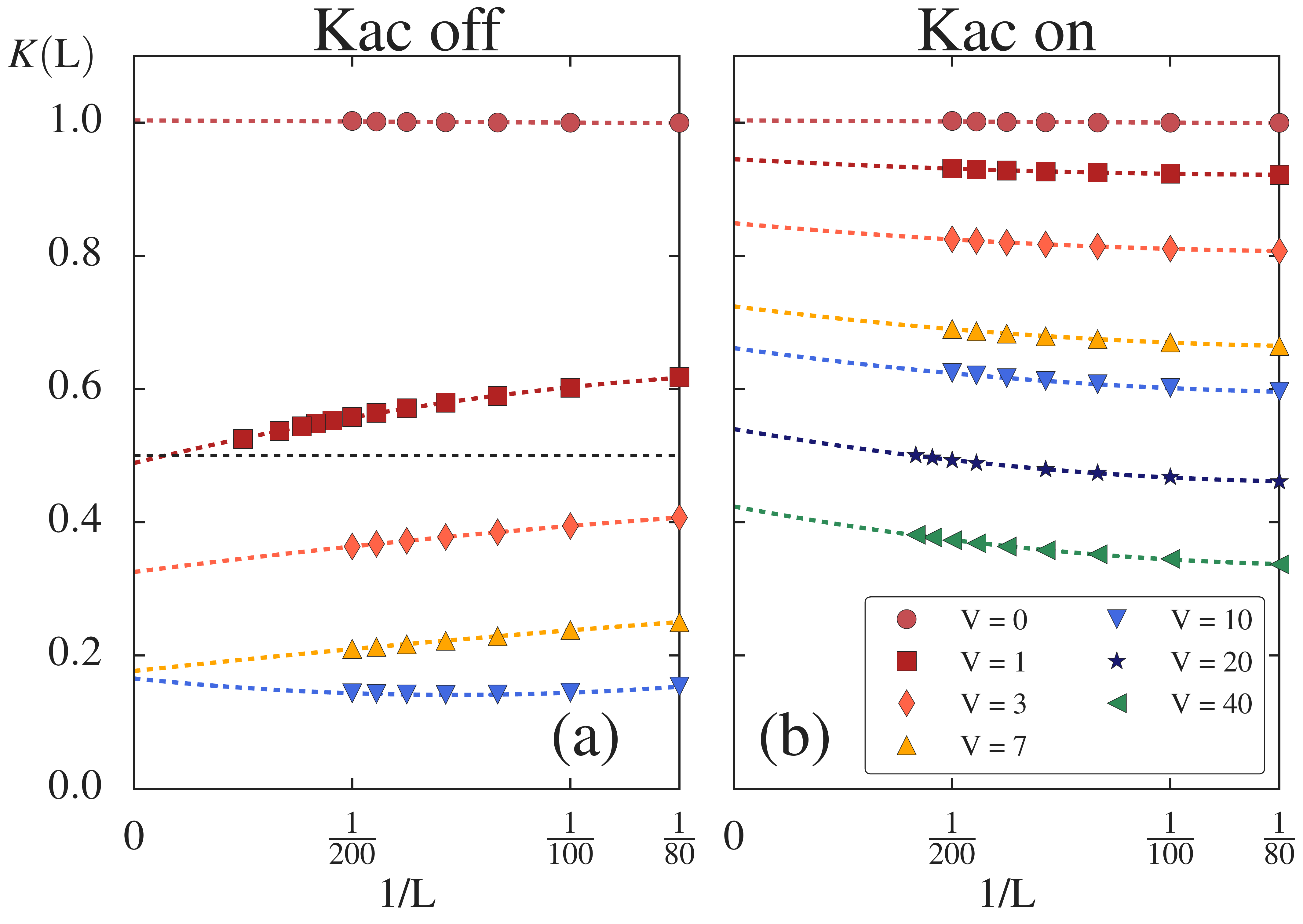}
\caption{Luttinger parameter $K$ computed numerically at half-filling for $\alpha=0.5$ and different $V$, by fitting the correlation function $\langle a_i^{\dagger} a_j \rangle$~\cite{Note2}. The critical value $K_c=0.5$ indicating the metal-insulator transition with nearest neighbor interaction is displayed as a black dashed line. In the absence of Kac's rescaling \textbf{(a)}, $K$ decreases when increasing $L$, lying below the critical line for $L\to \infty$ (insulating phase). In contrast, $K$ increases with $L$ in the presence of Kac's rescaling \textbf{(b)}, and remains finite even for very large $V$ (metallic phase). Extrapolation in the thermodynamic limit is obtained by fitting the data with the same function as in Fig.~\ref{Fig1} (dotted lines).}
\label{Fig2}
\end{figure}

\pagebreak

In order to gain further insights, we first compute the Luttinger parameter $K$ by fitting the correlation function $\langle a_i^{\dagger} a_j \rangle$, and represent it in Fig.~\ref{Fig2} for different $V$ and $\alpha=0.5$. We observe two opposite trends depending on whether Kac's rescaling is present or not. In the latter case [Fig.~\ref{Fig2} \textbf{(a)}], $K$ decreases when increasing $L$ and lies below the critical value $K_{c}=0.5$ for $L \to \infty$, which indicates an insulating phase. The case with Kac's rescaling is shown in Fig.~\ref{Fig2} \textbf{(b)}, where a finite $K$ is found for all $V$ in the thermodynamic limit. 

\begin{figure}[h]
\includegraphics[width=\columnwidth]{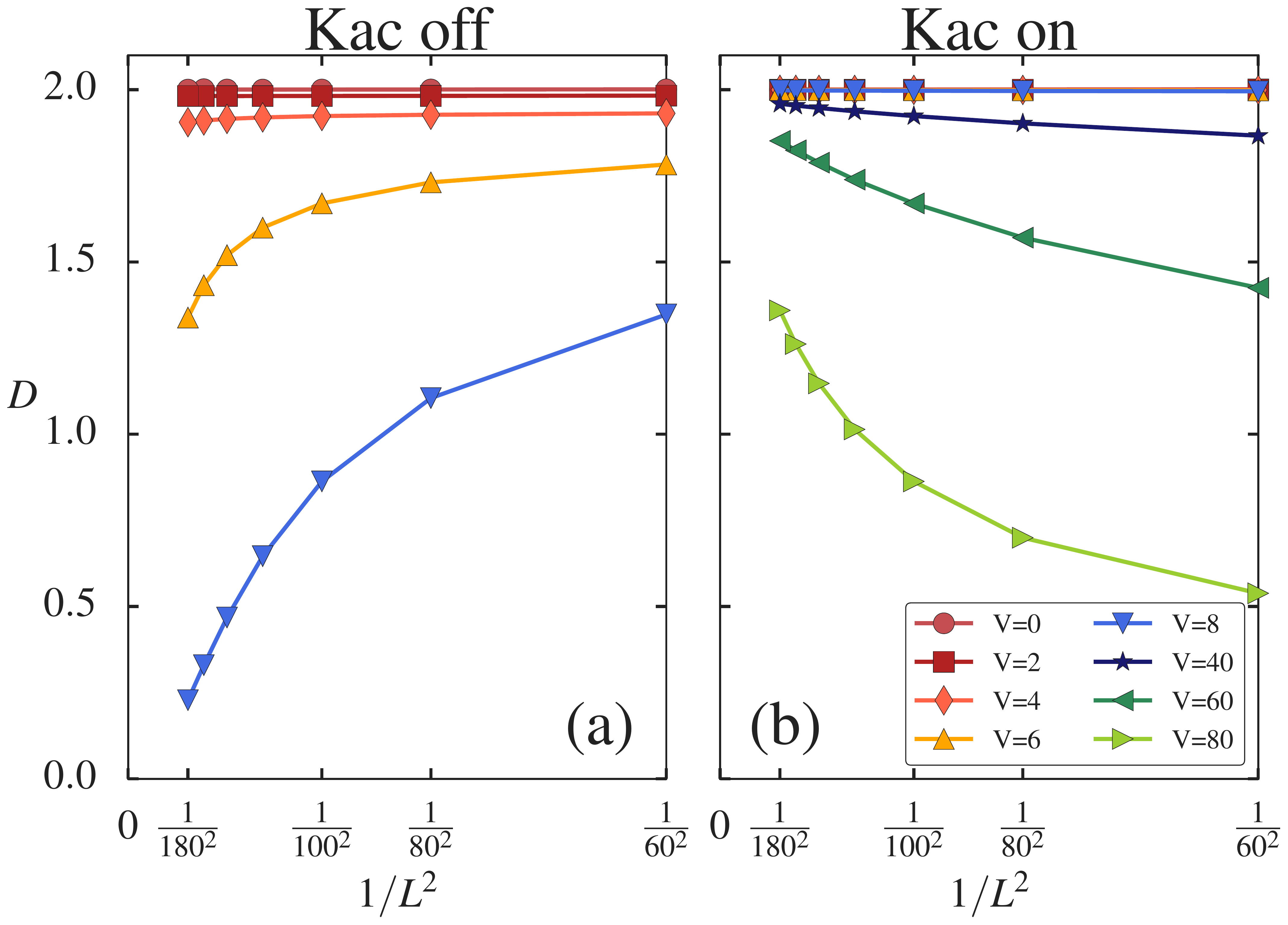}
\caption{Charge stiffness $D$ computed numerically from Eq.~(\ref{charge_stiff}) at half-filling, as a function of $1/L^{2}$ for $\alpha=0.5$ and different $V$. The magnetic flux $\Phi$ is implemented via the twisted boundary condition $c_1 = e^{i\Phi} c_{L+1}$~\cite{Schmitteckert2004}. Two opposite trends are observed depending on whether Kac's rescaling is present \textbf{(b)} or not \textbf{(a)}. While $D \to 0$ for $L\to \infty$ in the latter case (insulator), $D$ remains finite in the former case (metal).}
\label{Fig3}
\end{figure}

We then compute the charge stiffness~\cite{Giamarchi2004}
\begin{equation}
D =  \pi L \left\rvert \frac{\partial^{2} E_0 (\Phi)}{\partial \Phi^{2}} \right\rvert_{\Phi = 0},
\label{charge_stiff}
\end{equation}
which is proportional to the Drude weight~\cite{Kohn1964} and therefore provides valuable information on the metallic or insulating properties of the system. Moreover, it also gives a direct measure of the umklapp scattering strength. A large $D$ corresponds to a good metal, while an insulating phase features $D = 0$. The charge stiffness is computed numerically from the ground state energy $E_0$ by threading a flux $\Phi$ through the circular chain, and represented in Fig.~\ref{Fig3} as a function of $1/L^{2}$ for $\alpha=0.5$. In the absence of Kac's rescaling [Fig.~\ref{Fig3} \textbf{(a)}], $D$ decreases when increasing $L$ for any finite $V$. The latter drives the system towards an insulating phase ($D \to 0$) in the thermodynamic limit. In contrast, $D$ increases with $L$ in the presence of Kac's rescaling [Fig.~\ref{Fig3} \textbf{(b)}], which confirms the existence of a metallic behavior. In the thermodynamic limit, we find that $D \approx v_{\rm F}$ even for very large $V$, in surprisingly good agreement with the LL prediction $D=u K$ and Eq.~(\ref{eq: LL_parameters}). Note that we have performed a full numerical study showing that the conclusions drawn from Figs.~\ref{Fig2} and \ref{Fig3} can be unambiguously extended to the whole range $0\leq \alpha \leq 1$. 

\begin{figure}[ht]
\includegraphics[width=\columnwidth]{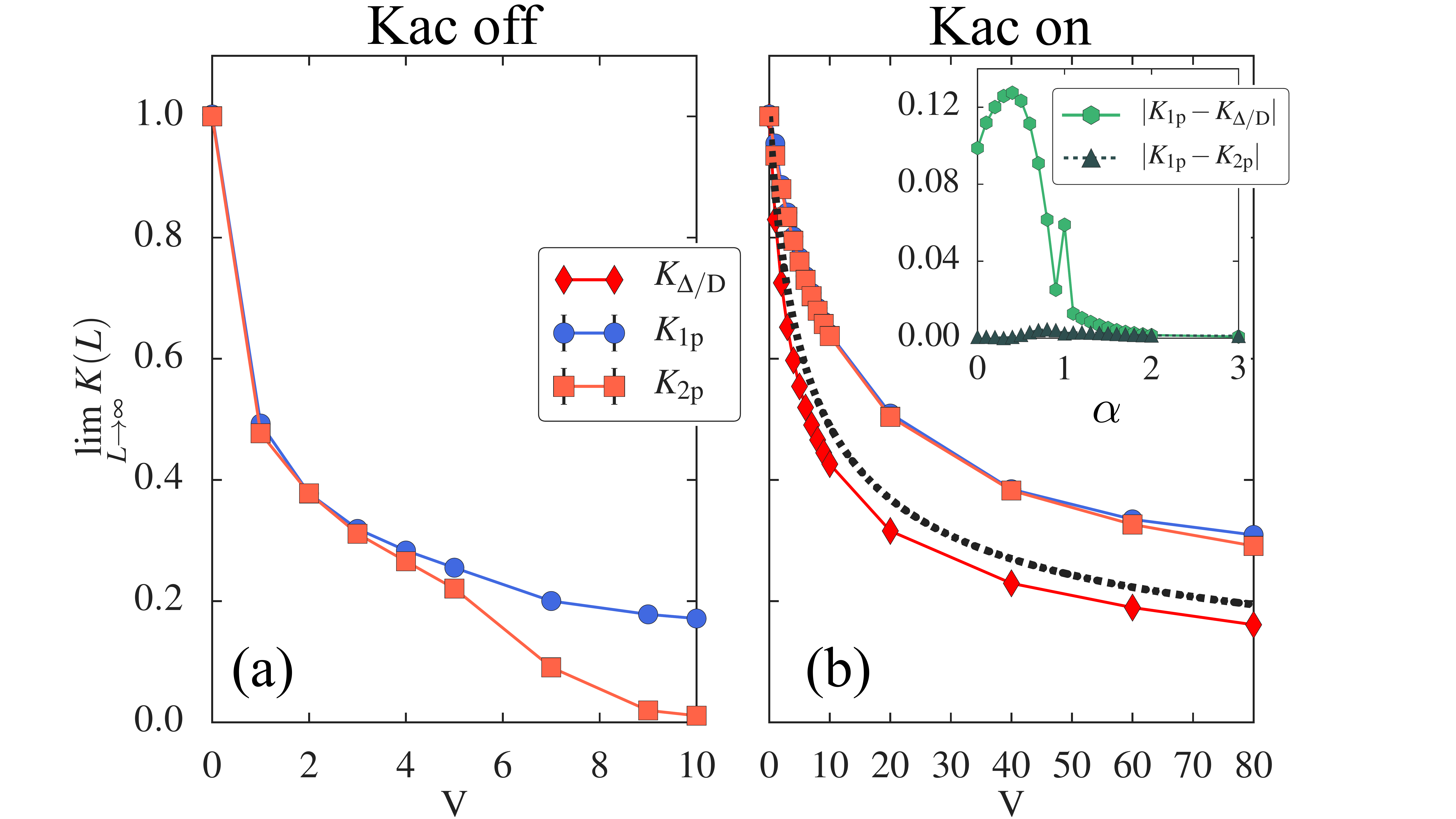}
\caption{Luttinger parameter $K$ extrapolated in the thermodynamic limit versus $V$ at half-filling and for $\alpha=0.5$, without \textbf{(a)} and with \textbf{(b)} Kac's rescaling. $K$ is computed in $3$ different ways: From the single-particle correlations ($K_{\rm 1p}$), the structure factor ($K_{\rm 2p}$), and from the gap and the charge stiffness ($K_{\rm \Delta/D}$). The formula obtained from Eq.~(\ref{eq: LL_parameters}) is displayed as a dotted line. Inset: $\vert K_{\rm 1p}-K_{\rm 2p}\vert$ and $\vert K_{\rm 1p}-K_{\rm \Delta/D} \vert$ versus $\alpha$ for $V=1.5$. A discrepancy between $K_{\rm 1p}$ and $K_{\rm 2p}$ is observed without Kac's rescaling, which indicates the breakdown of the LL theory (insulator). In contrast, the property $K_{\rm 1p}=K_{\rm 2p}$ observed with Kac's rescaling even for large $V$ suggests a metallic phase, which is not captured by the conventional LL theory since $K_{\rm \Delta/D}$ does not match $K_{\rm 1p,2p}$ for $\alpha <1$. In the short-range case $\alpha \gg 1$, one recovers $K_{\rm 1p,2p}=K_{\rm \Delta/D}$ in agreement with the standard LL theory.}
\label{Fig4}
\end{figure}

Now that we have demonstrated the metallic character of the ground state, we check the validity of the LL theory in Fig.~\ref{Fig4} by computing the parameter $K$ for $L \to \infty$ and $\alpha=0.5$ in three different ways: From the single-particle correlation function $\langle a^{\dagger}_{i} a_{j}\rangle$ (see above), from the static structure factor
\begin{equation}
S (q) = \frac{1}{L} \sum_{i,j} e^{\mi q \vert i-j \vert} \left(\langle n_{i} n_{j} \rangle - \langle n_{i} \rangle \langle n_{j} \rangle \right) 
\label{str_factor}
\end{equation}
as $K= L S(q=2 \pi /L)$, and from the relations $\pi \frac{u}{K}=\frac{\partial \Delta}{\partial (1/L)}$ and $u K=D$ stemming from the LL theory~\cite{Giamarchi2004}. In the absence of Kac's rescaling [Fig.~\ref{Fig4} \textbf{(a)}], a discrepancy between the values of $K$ extracted from the two correlation functions (labeled $K_{\rm 1p}$ and $K_{\rm 2p}$ in the figure) is observed, which indicates the breakdown of the LL theory related to the opening of a gap (insulating phase). The agreement obtained for small $V$ is attributed to the metal-like character at finite $L$ consistently with the data shown in Fig.~\ref{Fig2}. In the presence of Kac's rescaling [Fig.~\ref{Fig4} \textbf{(b)}], $K_{\rm 1p}$ and $K_{\rm 2p}$ match well up to very large $V$, while they match neither the formula $K=1/\sqrt{1+V/[\pi v_{\rm F} (1-\alpha)]}$ (dotted line) stemming from Eq.~(\ref{eq: LL_parameters}) nor $K$ obtained from $\Delta$ and $D$ (labeled $K_{\rm \Delta/D}$ in the figure). We find that this discrepancy holds in the entire range $0 \leq \alpha \leq 1$ (see inset), implying a breakdown of LL theory in the strong LR regime. For $\alpha >0$, this breakdown is only partial since numerics indicate that both $K_{\rm \Delta/D}$ and $K_{\rm 1p,2p}$ maintain the functional form 
\begin{equation}
K=\frac{1}{\sqrt{1+\gamma V/(\pi v_{\rm F})}}
\label{fun_form}
\end{equation}
with $\gamma$ finite for all $V$ [see Fig.~\ref{Fig4} \textbf{(b)}]. This is, however, not true for $\alpha=0$ which can be solved analytically~\cite{Note1}. In this case, the free fermion result $K_{\rm 1p,2p}=1$ for all $V$ extracted from the correlation functions differs significantly from the functional form Eq.~(\ref{fun_form}) ($\gamma=0$). Interestingly, we find that $K_{\rm \Delta/D}$ is correctly described by Eq.~(\ref{fun_form}) with $\gamma=1$, which corresponds exactly to the analytic prediction obtained from Eq.~(\ref{eq: LL_parameters}). Note that in the short-range case $\alpha \gg 1$, all methods predict the same $K$ as expected from the LL theory (see inset). The demonstration of this gapless, critical~\cite{Note1} metallic phase that does not fall into the conventional LL theory is a central result of this work. While here we focused on the case of half-filling, this phase appears in fact for all densities.  

In the supplemental material, we show that the long-wavelength excitations of this metallic phase have a linear dispersion, similar to a LL with short-range interactions. This can be readily shown by considering the continuous Hamiltonian Eq.~(\ref{cont_H}) with interaction potential $V^{(\alpha)}(x)=V/\left(x^{2}+a^{2}\right)^{\alpha/2}$. The diagonalization of this Hamiltonian in Fourier space provides the plasmon dispersion relation $\omega (q)= v_{\rm F} q \sqrt{1+V^{(\alpha)} (q)/(\pi v_{\rm F})}$. Kac's rescaling eliminates the long-wavelength divergence of the Fourier component $V^{(\alpha)} (q \to 0)$ and therefore leads to the dispersion relation of a metal with short-range interactions $\omega (q)\sim v_{\rm F}q$. Note that since the algebraic character of the interaction potential is preserved when using the Kac's prescription, the latter is thus ``weaker'' than Thomas-Fermi screening which turns the LR interaction into a short-range one.
  
 
In conclusion, we have shown that the low-energy properties of 1D hard-core bosons interacting via a LR potential are fundamentally modified when applying the Kac's prescription~\cite{Note4}, which restores energy extensivity and a well-defined thermodynamic limit. We find that the linear excitation spectrum of our unconventional metallic phase is also present for $d>1$ in the case $\alpha=1$. It is therefore an interesting prospect to investigate the properties of such an unconventional liquid with restored energy extensivity in higher dimensions~\cite{Tupitsyn2017}. Since cavity-mediated two-body interactions are naturally LR and extensive~\cite{Ritsch2013}, another perspective of this work is the exploration of the non-trivial thermodynamics of cold atoms in cavity-QED. For instance, the case $\alpha=0$ can be typically obtained when the spatial extent of the atomic cloud is much smaller than the cavity wavelength~\cite{Schutz2014}. 

\vspace{2mm}

\footnotetext[1]{See Supplemental Material including a sketch of the system, an example of critical (algebraic decay) single-particle correlation function, a discussion on the suppression of the plasmon mode when restoring energy extensivity, and the exact solution of the model for $\alpha=0$. The Supplemental Material includes Refs.~\cite{Giamarchi2004,Schulz1993,Cazalilla2004,Batrouni2005}.}

\footnotetext[2]{We use the fitting function derived in Ref.~\cite{Cazalilla2004} from conformal field theory with periodic boudary conditions on a ring.}

\footnotetext[3]{This phase was refered to as a strange metal in Ref.~\cite{Li2019}, since a finite gap $\Delta = V$ is found for $L\to \infty$ in the absence of Kac's rescaling. Upon restoring a well-defined thermodynamic limit ($V \to V/L$), we find that the gap $\Delta \sim (V + 2\pi t)/L \to 0$ for $L\to \infty$ consistently with a metallic phase.}

\footnotetext[4]{It is worth commenting on an alternative Kac's prescription, which consists in dividing the whole Hamiltonian Eq.~(\ref{full_H}) by $\Lambda_{\alpha} (L)$. The behaviors of the correlation function and of the charge stiffness are not affected by such a global rescaling, and are therefore those of an insulator. However, the charge gap vanishes in the thermodynamic limit since the energy of the ground state $E_{0}$ is simply replaced by $E_{0}/L$, and the system exhibits an hybrid insulating/metallic behavior.}

{\textbf{Acknowledgements} --- }
We are grateful to S.~Ruffo, S.~Sch\"{u}tz, and D.~Vodola for stimulating discussions. Work in Strasbourg was supported by the ANR - ``ERA-NET QuantERA'' - Projet ``RouTe'' (ANR-18-QUAN-0005-01), and LabEx NIE. G.~P. Acknowledges support from the Institut Universitaire de France (IUF) and USIAS. G. M. was also supported by the French National Research Agency (ANR) through the ``Programme d'Investissement d'Avenir'' under contract ANR-17-EURE-0024. Computing time was provided by the HPC-UdS. N.~D.~acknowledges financial support by Deutsche Forschungsgemeinschaft (DFG) via Collaborative Research Centre SFB 1225 (ISOQUANT) and under Germanys Excellence Strategy EXC-2181/1-390900948 (Heidelberg STRUCTURES Excellence Cluster).

\bibliographystyle{apsrev4-1}
\bibliography{biblio_main}

\clearpage

\onecolumngrid

\begin{center}
	\textbf{\large Effects of energy extensivity on the quantum phases of long-range interacting systems\\ [.3cm] -- Supplemental Material --}\\[.4cm]
	T.~Botzung,$^{1}$ D.~Hagenm\"uller,$^{1}$ G.~Masella,$^{1}$ J.~Dubail,$^{2}$ N.~Defenu,$^{3}$ A.~Trombettoni,$^{4}$ and G.~Pupillo$^{1}$\\[.1cm]
	{\itshape ${}^1$ISIS (UMR 7006) and icFRC, University of Strasbourg and CNRS, 67000 Strasbourg, France\\
		${}^2$LPCT (UMR7019), Universit\'{e} de Lorraine and CNRS, F-54506 Vandoeuvre-les-Nancy, France\\
		${}^3$Institute for Theoretical Physics, Heidelberg University, D-69120 Heidelberg, Germany\\
		${}^4$SISSA and INFN, Sezione di Trieste, I-34136 Trieste, Italy\\}
	(Dated: \today)\\[1cm]
\end{center}

\setcounter{equation}{0}
\setcounter{figure}{0}
\setcounter{table}{0}
\setcounter{page}{1}
\renewcommand{\theequation}{S\arabic{equation}}
\renewcommand{\thefigure}{S\arabic{figure}}

In Sec.~\ref{exten}, we show that the plasmon mode of a $d$-dimensional system of long-range (LR) interacting fermions (or hard-core bosons) is suppressed when using the Kac prescription in the strong long-range regime $\alpha \leq d$. In Sec.~\ref{alpha0}, we solve the extreme case $\alpha=0$ exactly using a mean-field approach showing that the system is in a free fermion phase with a charge gap $\Delta \sim (V+2\pi t)/L \to 0$ for $L \to \infty$, which is not described by the Luttinger liquid theory. 

\begin{figure}[ht]
\includegraphics[width=0.7\columnwidth]{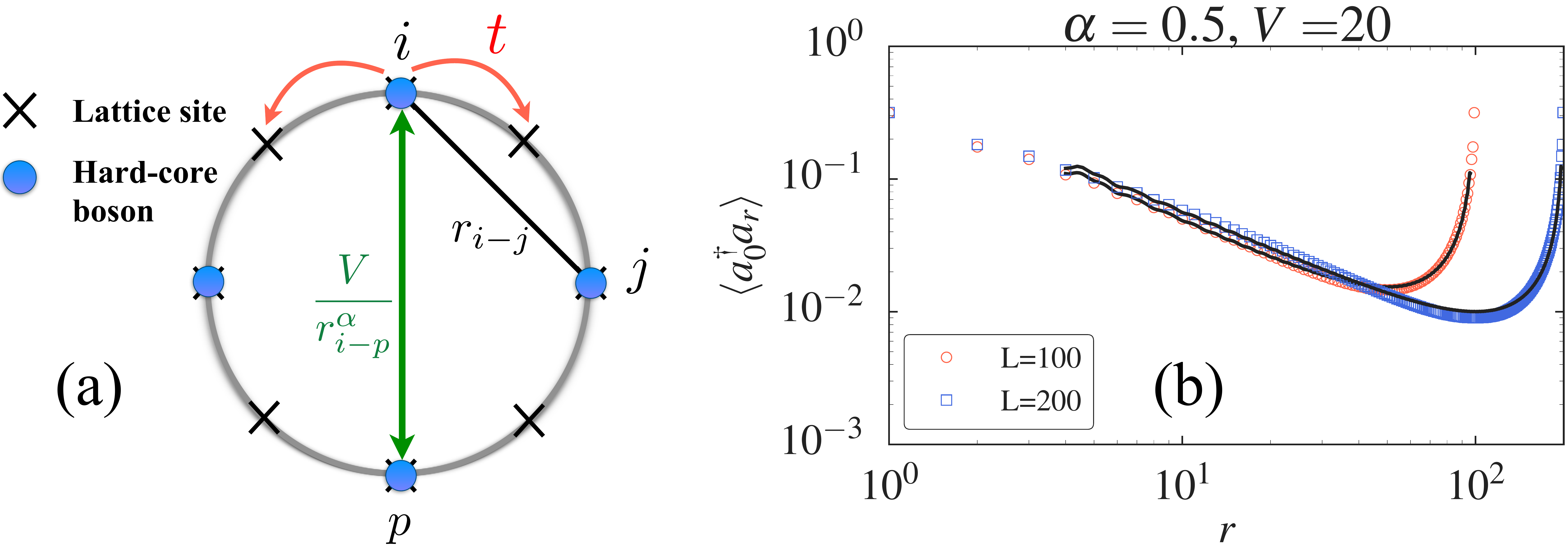}
\caption{\textbf{(a)} Sketch of the physical system described by the Hamiltonian Eq.~(1) of the main text, consisting of a circular chain (periodic boundary conditions). \textbf{(b)} Critical single-particle correlation function $\langle a^{\dagger}_{0} a_{r} \rangle$ at half-filling, for $\alpha=0.5$, $V=20$, and two different system sizes $L=100,200$. The numerical data is fitted with the function described in Ref.~[\onlinecite{Cazalilla2004}] for a circular chain (black lines).}
\label{Fig1_sup}
\end{figure}

\section{Effect of energy extensivity on plasmon modes}
\label{exten}

\subsection{One-dimensional Luttinger liquid}

\begin{figure}[h]
\includegraphics[width=0.6\columnwidth]{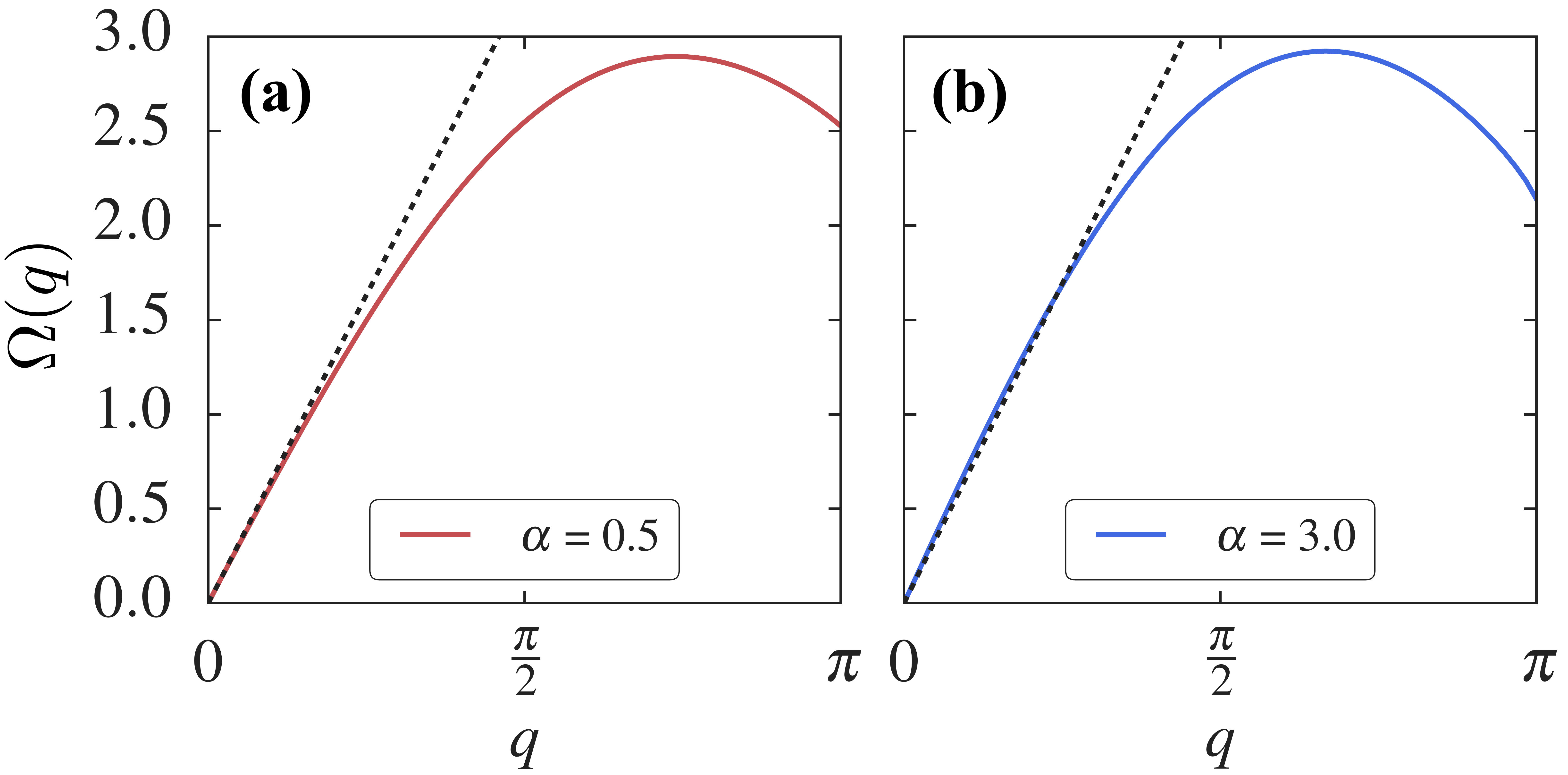}
\caption{Upper bound of the excitation spectrum $\Omega({q})$ (in units of $t$) in the Feynman approximation (colored lines) computed at half-filling for $V=0.5$ in the strong LR regime $\alpha=0.5$ \textbf{(a)}, and in the short-range case $\alpha=3$ \textbf{(b)}. The dispersion relation Eq.~(\ref{plat}) after Kac's rescaling, namely $\omega(q)=v_{\rm F}q\sqrt{1+V/\pi}$ for $\alpha=0.5$ and $\omega(q)=v_{\rm F}q\sqrt{1+V/(\sqrt{\pi} \Gamma(3/2) v_{\rm F})}$ for $\alpha=3$ is represented as a black dotted line in the long-wavelength regime $q\to 0$. The proximity of the Mott transition ($V=2$ for $\alpha \to \infty$) in the short-range case is responsible for the more pronounced minimum at $q=\pi$ (charge density wave).}
\label{Fig3_sup}
\end{figure}

We consider a one-dimensional system of length $L$ containing $N$ fermions interacting via the LR potential 
\begin{equation*}
V^{\alpha}(x) = \frac{V}{\left(x^{2}+a^{2}\right)^{\alpha/2}} \qquad 0<\alpha \leq 1. 
\end{equation*}
Here, $a$ denotes a short-distance cutoff that can be identified with, e.g., the lattice spacing. In the vicinity of the Fermi level, the low-energy Hamiltonian can be decomposed into the contributions of left (L) and right (R) movers as 
\begin{equation}
H = \sum_{k}\sum_{r={\rm L},{\rm R}} \hbar v_{\rm F} \left(\eta_{r} k-k_{\rm F}\right) c^{\dagger}_{r,k} c_{r,k} + \frac{1}{2}\int \! {\rm d} x {\rm d} x' \rho(x) V^{\alpha} (x-x') \rho(x'),
\label{hamol}
\end{equation}  
where $v_{\rm F}$ is the Fermi velocity, $k_{\rm F}$ the Fermi wave vector, $\eta_{\rm R}=+1$, $\eta_{\rm L}=-1$, and $\rho (x) = \sum_{r={\rm L},{\rm R}} \rho_{r} (x)$ with $\rho_{r} (x) = \frac{1}{L} \sum_{k,q} e^{i q x} c^{\dagger}_{r,k+q} c_{r,k}$. Bosonization assumes that the low-energy properties of the Hamiltonian Eq.~(\ref{hamol}) are governed by the long-wavelength fluctuations of the density $\rho (x)$. Using the standard techniques described in Ref.~[\onlinecite{Giamarchi2004}], $H$ can be approximately written (for $L \to \infty$) in the quadratic form
\begin{equation*}
\label{Hamiltonian_F}
H = \frac{1}{2\pi} \sum_q u(q) K(q) \pi^{2} \Pi (q) \Pi (-q) + \frac{u(q)}{K (q)} q^{2} \phi(q)\phi(-q),
\end{equation*}
where $u(q)$ denotes the velocity of the excitations and $K(q)$ is the Luttinger parameter governing the decay of correlations at long distances. The latter satisfy the relations 
\begin{align}
\text{\hspace{15mm}} u(q) K(q) = v_{\rm F} \text{\hspace{20mm}}
\frac{u(q)}{K(q)} =  v_{\rm F}\left[ 1 + \frac{V^{(\alpha)} (q)}{\pi v_{\rm F}} \right].
\label{lutt_par}
\end{align}
The Fourier transform of the interaction potential reads
\begin{equation}
V^{\alpha}(q) = \int \! {\rm d}x \, V^{\alpha}(x) e^{-i q x}= V \frac{2\sqrt{\pi}}{\Gamma\left(\frac{\alpha}{2}\right) 2^{\frac{\alpha - 1}{2}}}\left(\frac{\vert q \vert}{a}\right)^{\frac{\alpha - 1}{2}} \mathcal{K}_{\frac{\alpha - 1}{2}} (a \vert q\vert),
\label{poterie}
\end{equation}
and the two fields $\Pi (q)=\int \! {\rm d}x \, \Pi (x) e^{-i q x}$ and $\phi(q) =\int \! {\rm d}x \, \phi(x) e^{-i q x}$ are the Fourier transforms of the canonically conjugate fields $\Pi(x)=\frac{1}{\pi}{\bm \nabla} \theta (x)$ and $\phi(x)$ with
\begin{align*}
\phi(x) &= - (N_{\rm R}+N_{\rm L}) \frac{\pi x}{L} - \frac{\mi \pi}{L} \sum_{q\neq 0} \frac{1}{q} e^{-\beta \vert q \vert/2 - \mi q x} \left(\rho_{\rm R} (q) + \rho_{\rm L} (q)\right) \\
 \theta (x) &= (N_{\rm R}-N_{\rm L}) \frac{\pi x}{L} + \frac{\mi \pi}{L} \sum_{q\neq 0} \frac{1}{q} e^{-\beta \vert q \vert/2 - \mi q x} \left(\rho_{\rm R} (q) - \rho_{\rm L} (q)\right).
\end{align*}
Here, $\beta$ is a (small) cutoff regularizing the theory, $N_{r}=\sum_{k} c^{\dagger}_{r,k}c_{r,k}- \langle c^{\dagger}_{r,k}c_{r,k} \rangle$, and $\rho_{r} (q)=\sum_{k} c^{\dagger}_{r,k+q}c_{r,k}$. The plasmon dispersion relation follows from Eq.~(\ref{lutt_par}) and reads
\begin{equation}
\omega (q)= u(q) \vert q\vert = v_{\rm F} \vert q \vert \sqrt{1+\frac{V^{(\alpha)} (q)}{\pi v_{\rm F}}}. 
\label{plat}
\end{equation}
The potential Eq.~(\ref{poterie}) exhibits a long-wavelength divergence ($q \to 0$), namely $V^{\alpha} (q)\sim \vert q \vert^{\alpha -1}$ for $0<\alpha <1$ and $V^{\alpha} (q)\sim \log \vert q \vert$ for $\alpha=1$. In the latter case, Eq.~(\ref{plat}) provides the 1D plasmon dispersion $\omega (q)\sim \vert q \vert \sqrt{\log \vert q \vert}$ stemming from Coulomb interactions~\cite{Schulz1993}. When rescaling the interaction potential by the Kac's factor $\Lambda_{\alpha} (L)=L^{1-\alpha}$ for $0\leq\alpha <1$ and $\Lambda_{\alpha} (L)=\log (L)$ for $\alpha =1$, it is easy to check that the long-wavelength divergence of the potential is removed by considering the limit $q =\frac{2\pi}{L} \to 0$. As a consequence, one recovers the sound wave dispersion relation $\omega (q)\sim \vert q \vert$ of a metal with short-range interactions. This result is confirmed by looking at the upper bound of the excitation spectrum $\Omega({q})=E (q)/S(q)$ in the Feynman approximation~\cite{Batrouni2005} represented in Fig.~\ref{Fig3_sup}, where $E(q)=(t/L) \left[1-\cos (q) \right] \langle \sum_{i} a^{\dagger}_{i} a_{i+1} + \textrm{h.c.} \rangle$ and $S(q)$ is the structure factor defined by Eq.~(6) of the main text. 

\subsection{Generalization to higher dimensions}

This result can be easily generalized to higher dimensions $d =2,3$ by looking at the zeros of the dielectric function in the framework of the random phase approximation (RPA):
\begin{equation}
\epsilon ({\bf q},\omega)= 1 - \chi ({\bf q},\omega) V^{\alpha} ({\bf q}) =0,
\label{die}
\end{equation}
where 
\begin{equation*}
\chi ({\bf q},\omega)= \frac{1}{\mathcal{V}} \sum_{\bf k} \frac{n_{\bf k} - n_{{\bf k}+{\bf q}}}{\hbar\omega + E_{\bf k}- E_{{\bf k}+{\bf q}} + \mi \eta}
\end{equation*}  
denotes the one-spin density-density response function (Lindhard function), $\mathcal{V}$ the volume, and $n_{\bf k}$ the occupation number of a state with wave vector ${\bf k}$ and energy $E_{\bf k}=\frac{\hbar^{2} \vert {\bf k} \vert^{2}}{2 m}$ ($m$ is the particle mass). For $\alpha=1$, the Fourier transform of the Coulomb potential is
\begin{alignat}{3}
&V^{1} (q)&&\sim \log \vert q \vert \qquad && d=1 \nonumber \\
&V^{1} ({\bf q})&&\sim \frac{1}{\vert {\bf q} \vert} \qquad && d=2 \nonumber \\
&V^{1} ({\bf q})&&\sim \frac{1}{\vert {\bf q} \vert^{2}} \qquad && d=3.
\label{poRPA}
\end{alignat}  
In the dynamical limit $\omega \gg \vert {\bf q} \vert v_{\rm F}$, the Lindhard function can be approximated by $\chi ({\bf q},\omega) = \frac{\rho_{0} \vert {\bf q} \vert^{2}}{m \omega^{2}}$ with $\rho_{0}$ the average fermion density. Using this expression together with Eq.~(\ref{poRPA}) into Eq.~(\ref{die}), one finds the plasmon energies
\begin{alignat}{3}
&\omega &&\sim \vert q \vert \sqrt{\log \vert q \vert} \qquad && d=1 \nonumber \\
&\omega &&\sim \sqrt{\vert {\bf q}\vert} \qquad && d=2 \nonumber \\
& \omega &&\sim {\rm cst} \qquad && d=3.
\end{alignat}  
When using the Kac's prescription, namely dividing the potential by the factor $L^{d - 1}$, the long-wavelength divergence ($q=\frac{2\pi}{L}\to 0$) is removed and one recovers the sound wave dispersion relation $\omega \sim \vert {\bf q} \vert$ for $d=1,2,3$.   

\section{The extreme case $\alpha=0$}
\label{alpha0}

The particular case $\alpha=0$ can be solved exactly using a mean field approach on the Hamiltonian Eq.~(1) of the main text. For convenience, we use fermions instead of hard-core bosons since the results are equivalent in both cases. We start from the Hamiltonian
\begin{align*}
H= -t \sum_{i=0}^{L-1} \left(c^{\dagger}_i  c_{i+1}  + {\rm h.c.} \right) + \frac{V}{2} \sum_{i\neq j} n_{i} n_{j},
\end{align*}
where $c_i$ ($c_i^{\dagger}$) annihilates (creates) a fermion on site $i=1,\cdots,L$, and $n_i = c_i^{\dagger} c_i$ is the local density. Writting the density-density interaction as $n_{i} n_{j}\approx n_{i} \langle n_{j}\rangle + n_{j} \langle n_{i}\rangle - \langle n_{i}\rangle \langle n_{j}\rangle$, the mean-field Hamiltonian reads
\begin{equation}
H_{\rm mf}= \sum_{k} \left[\left(N-1\right) V - 2 t\cos \left(k \right) \right] c^{\dagger}_{k} c_{k} - \frac{N \left(N-1 \right)V}{2}, 
\label{mfi}
\end{equation}  
with $N=L/2$ the number of fermions (at half-filling), and $c_{k}=\frac{1}{\sqrt{L}} \sum_{j=0}^{L-1} c_{j} e^{-2\mi \pi k j/L}$. The energy of the ground state for, e.g., $N$ even, is derived from Eq.~(\ref{mfi}) with anti-periodic boundary conditions as 
\begin{equation*}
E_{0} (N)=\frac{N \left(N-1 \right)V}{2} - 2 t \sum_{k=-L/4}^{L/4-1} \cos \left[\frac{2\pi k}{L}+ \left(\frac{\pi}{L}\right) \right] =\frac{N \left(N-1 \right)V}{2} -2 t \csc \left(\frac{\pi}{L}\right).
\end{equation*}
One then has to consider periodic boundary conditions for $N \pm 1$ fermions, which leads to
\begin{align*}
E_{0} (N+1)&=\frac{\left(N+1 \right)N V}{2} - 2 t \sum_{k=-L/4}^{L/4} \cos \left(\frac{2\pi k}{L} \right) = \frac{\left(N+1 \right)N V}{2} -2 t \cot \left(\frac{\pi}{L}\right) \\
E_{0} (N-1)&=\frac{\left(N-1 \right) \left(N -2 \right)V}{2} - 2 t \sum_{k=-L/4 + 1}^{L/4 - 1} \cos \left(\frac{2\pi k}{L} \right) = \frac{\left(N-1 \right) \left(N -2 \right)V}{2} -2 t \cot \left(\frac{\pi}{L}\right).
\end{align*}
The charge gap thus reads $\Delta\equiv E_{0} (N+1) + E_{0} (N-1) - 2 E_{0} (N) = V + 4 t \tan \left(\frac{\pi}{2L} \right)$, and becomes $\Delta\sim \left(V+2\pi t\right)/L \to 0$ for $L\to \infty$ when using the Kac's prescription $V \to V/L$. The Luttinger parameters $u/K$ and $u K$ can be related to the first derivative of the single-particle charge gap as $\frac{\partial \Delta}{\partial (1/L)}=\pi \frac{u}{K}$, and to the charge stiffness~\cite{Giamarchi2004} as
\begin{equation*}
D = \pi L \left\rvert \frac{\partial^{2} E_0 (N,\Phi)}{\partial \Phi^{2}} \right\rvert_{\Phi = 0} = u K.
\end{equation*}
Here, $\Phi =2\pi \phi/\phi_{0}$ denotes a flux threading the (circular) chain in units of the flux quantum $\phi_{0}=h/e$. This flux can be taken into account by multiplying the hopping energy by an Aharonov-Bohm phase phase $e^{\pm \mi \Phi/L}$ as
\begin{align*}
H (\Phi)= -t\sum_{i=0}^{L-1} \left( e^{\mi \Phi/L} c^{\dagger}_i  c_{i+1}  + {\rm h.c.} \right) + \frac{V}{2} \sum_{i\neq j} n_{i} n_{j}.
\end{align*}
The energy of the Hartree-Fock ground state is derived as 
\begin{equation*}
E_{0} (N,\Phi)=\frac{N \left(N-1 \right)V}{2} -2 t \csc \left(\frac{\pi}{L}\right) \cos \left(\frac{\Phi}{L} \right),
\end{equation*}
which provides $D = 2 t = v_{\rm F}$ for $L \to \infty$. The Luttinger parameters extracted from the charge gap and from the charge stiffness thus read 
\begin{align}
u K = v_{\rm F} \text{\hspace{15mm}}
\frac{u}{K} =  v_{\rm F}\left[ 1 + \frac{V}{\pi v_{\rm F}} \right],
\label{lutt_par2}
\end{align}
and coincide exactly with the analytic prediction Eq.~(3) of the main text. Since the mean-field Hamiltonian Eq.~(\ref{mfi}) corresponds to that of free fermions up to a constant shift $\propto V$, it is straightforward to calculate the Luttinger parameter $K$ from the single-particle correlation function
\begin{align}
\langle c^{\dagger}_{i} c_{j} \rangle = \frac{1}{L} \sum_{k} e^{\mi k (i-j)} n_{k} = \frac{1}{2 \mi \pi} \frac{e^{\mi k_{\rm F} (i-j)}}{i-j} \sim (i-j)^{-1},
\label{corr_1p}
\end{align}
and from the long-wavelength limit of the static structure factor
\begin{align}
S(q) \equiv \frac{1}{L} \sum_{i,j} e^{\mi q (i-j)} \left(\langle n_{i} n_{j} \rangle - \langle n_{i} \rangle \langle n_{j} \rangle \right) = \frac{1}{L} \sum_{k,k'} \left(\langle c^{\dagger}_{k} c_{k-q} c^{\dagger}_{k'} c_{k'+q} \rangle - \langle c^{\dagger}_{k} c_{k-q} \rangle \langle c^{\dagger}_{k'} c_{k'+q} \rangle \right) \to_{q\to 0} \frac{1}{L}. 
\label{corr_2p}
\end{align}
Note that the only non-vanishing contribution to the last equation stems from the term $\propto \langle c^{\dagger}_{k} c_{k'+q} \rangle \langle c_{k-q} c^{\dagger}_{k'} \rangle$, which is finite only at the two edges of the Fermi sea where $n_{k}=1/2$. Comparing Eqs.~(\ref{corr_1p}) and (\ref{corr_2p}) to the predictions $\langle c^{\dagger}_{i} c_{j} \rangle \sim (i-j)^{-\frac{K+(1/K)}{2}}$ (for fermions) and $K=L S(q \to 0)$ of the Luttinger liquid theory, we thus find $K=1$ for all $V$ in disagreement with the result $K=1/\sqrt{1+ V/(\pi v_{\rm F})}$ obtained from Eq.~(\ref{lutt_par2}). This suggests a breakdown of the Luttinger liquid theory in the extreme case $\alpha=0$.

\end{document}